\documentclass[reprint,
 amsmath,amssymb,
 aps,
]{revtex4-1}
\usepackage{graphicx}% Include figure files
\usepackage{dcolumn}% Align table columns on decimal point
\usepackage{bm}% bold math

\usepackage{siunitx}
\usepackage[disable]{todonotes} % notes not showed

\begin{document}

\preprint{APS/123-QED}

\title{Force distribution in a semiflexible loop}
% Force line breaks with \\

\author{James T. Waters}
% \altaffiliation[Also at ]{Physics Department, XYZ University.}
\author{Harold D. Kim}%
 \email{harold.kim@physics.gatech.edu}
\affiliation{%
 School of Physics, Georgia Institute of Technology\\
 837 State Street, Atlanta, GA 30332-0430
}%


\date{\today}% It is always \today, today,

\begin{abstract}
Loops undergoing thermal fluctuations are prevalent in nature. Ring-like or cross-linked polymers, cyclic macromolecules, and protein-mediated DNA loops all belong to this category. Stability of these molecules are generally described in terms of free energy, an average quantity, but it may also be impacted by local fluctuating forces acting within these systems. The full distribution of these forces can thus give us insights into mechanochemistry beyond the predictive capability of thermodynamics. In this paper, we study the force exerted by an inextensible semiflexible polymer constrained in a looped state. By using a novel simulation method termed ``phase-space sampling", we generate the equilibrium distribution of chain conformations in both position and momentum space. We compute the constraint forces between the two ends of the loop in this chain ensemble using Lagrangian mechanics, and show that the mean of these forces is equal to the thermodynamic force. By analyzing kinetic and potential contributions to the forces, we find that the mean force acts in the direction of increasing extension not because of bending stress, but in spite of it. Furthermore, we obtain a distribution of constraint forces as a function of chain length, extension, and stiffness. Notably, increasing contour length decreases the average force, but the additional freedom allows fluctuations in the constraint force to increase. The force distribution is asymmetric and falls off less sharply than a Gaussian distribution. Our work exemplifies a system where large-amplitude fluctuations occur in a way unforeseen by a purely thermodynamic framework, and offers novel computational tools useful for efficient, unbiased simulation of a constrained system. 
\end{abstract}

\maketitle

%\tableofcontents
\listoftodos
\onecolumngrid
\section{Introduction}
A looped state where two ends of a connected chain are tied to each other is a commonly occurring geometry in nature. The mechanics of loops at the macroscopic scale is straightforward to understand based on elasticity. However, loops at the molecular level constantly undergo fast, random thermal fluctuations. Ring-shaped or cross-linked polymers, cyclic macromolecules\cite{laurent2009synthetic,deffieux2009naturally}, and DNA loops\cite{wong2008interconvertible,le2014probing,lawrimore2015dna} are some common examples of molecular loops. Polymers whose ends are held at fixed points also belong to this category, which can be realized in single-molecule pulling experiments\cite{noy2011force,manca2014equivalence}. These loops all exert forces internally within bonds or externally between the ends. These forces may impact stability and reactivity of the loop itself\cite{akbulatov2012force} or be transmitted to other molecules joining its ends\cite{villa2005structural}. Hence, understanding the force profile in a loop geometry may have relevance in mechanochemistry\cite{craig2012mechanochemistry} and gene regulation\cite{saiz2012physics,cournac2013dna}. 

%paragraph on the technical difficulty

However, the force profile of a sizable loop of fixed end-to-end distance has not been extensively studied due to many technical challenges.
First, the effect of the heat bath in addition to the time- and length-scales of loop dynamics are often too enormous to cover by molecular dynamics simulations. Second, the widely popular inextensible semiflexible chain model known as the Kratky-Porod wormlike chain poses analytical difficulties in statistical mechanics due to the fixed bond length constraints. Third, because of these constraints, it is also computationally challenging to sample looped conformations in phase space in an unbiased manner.

In this paper, we investigate the force that is generated between the two ends of a looped chain held at constant extension. In this so-called isometric ensemble, the force conjugate to the fixed extension arises due to two separate mechanisms. Bending along the chain leads to an elastic restoring force, while thermally excited mass points of the chain give rise to an inertial force. We previously showed that for a freely-jointed chain fixed at one end, the mean of the inertial forces is equal to the entropic force derived from thermodynamics\cite{waters2015calculation}. To extend this mechanical analysis to the loop geometry, we introduce hierarchical coordinates that allow easy, direct sampling of closed chains both in position and momentum space while preserving all internal and external length constraints. To differentiate our method from conventional Monte Carlo methods that sample position space only, we term our simulation method ``phase-space sampling". Using this method, we obtained the isometric ensemble of looped microstates, from which we calculate the constraint force conjugate to the end-to-end distance. We show that a delicate balance between elastic and inertial forces with large amplitudes lead to a small mean constraint force equal to the generalized force predicted from statistical mechanics. The constraint forces follow a broad, asymmetric distribution with a width that increases with chain length. To the best of our knowledge, our study is the first to compute the force distribution of an inextensible semiflexible polymer in the isometric ensemble, which is not analytically tractable.

%We consider a semiflexible polymer with fixed end-to-end distance which is known as the isometric (Helmholtz) ensemble. This ensemble has been extensively studied theoretically because of its relevance to the force-displacement relationship that can be measured experimentally\cite{manca2014equivalence}. Here, we developed a novel set of generalized coordinates that can be used to construct the isometric ensemble for a semiflexible chain with rigid segments and fixed contour length (Kratky-Porod wormlike chain). We used a Monte Carlo simulation to sample the polymer microstates in the phase space, and 

\section{Methods}
\subsection{Overview of Phase-Space Sampling}
\begin{figure}[th!]
\includegraphics[width=15cm]{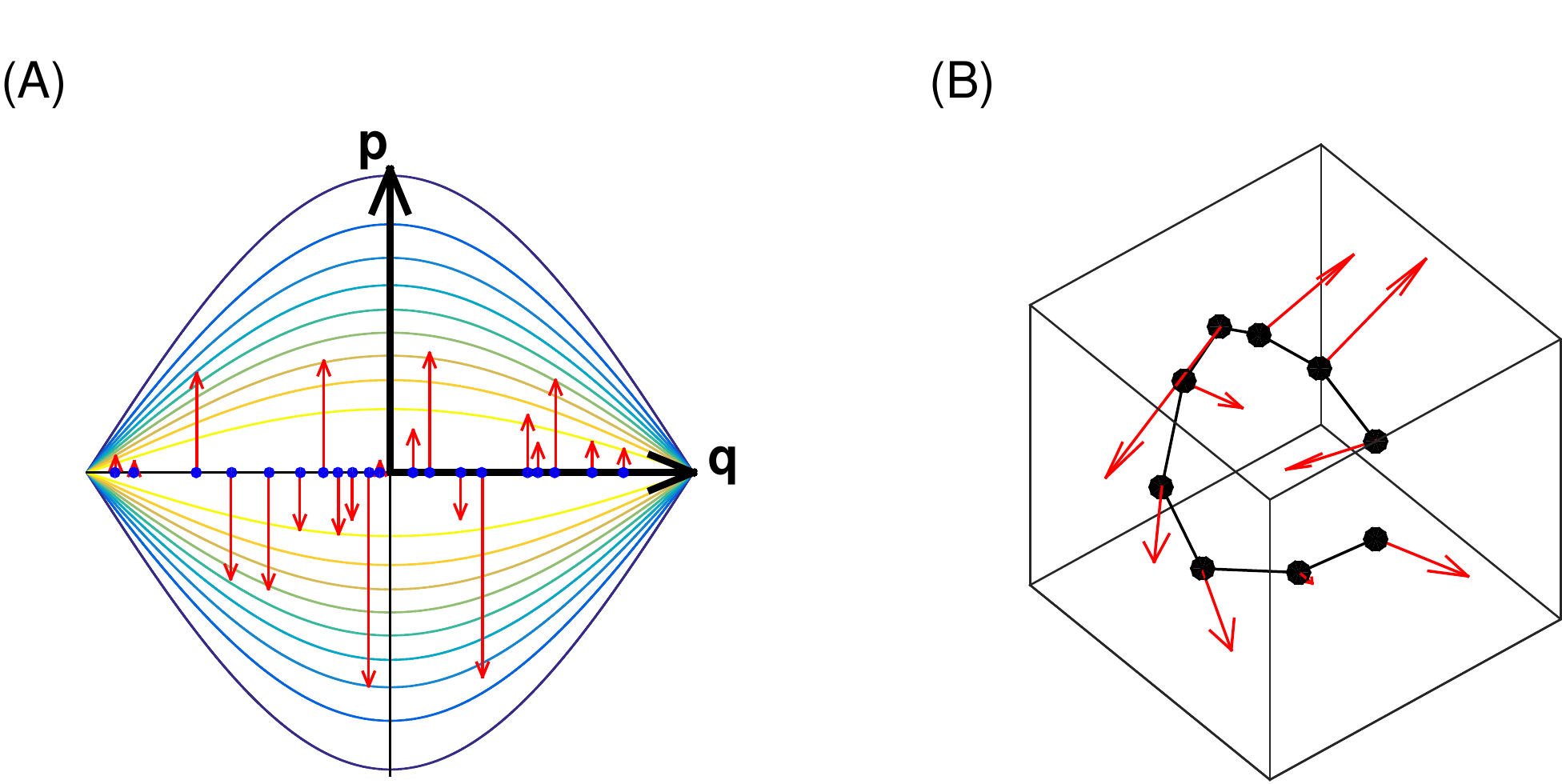}
\caption{General schematic of the phase-space sampling method. (A) The contour plot represents the probability density as a function of the position and momentum coordinates with the color indicating the magnitude of the density. As spatial coordinates $q_i$ vary, the size of the corresponding momentum space varies too. Conformations (blue dots) are chosen weighted by both the potential energy and the size of the momentum space at that point, obtained from the determinant of the metric tensor. After generating an ensemble of conformations, momenta (red arrows) are assigned to each one under the assumption the system is in a heat bath at some temperature. (B) Bead-rod representation of a sample loop configuration. Red arrows at mass points indicate velocities. A sample conformation from (A) is picked from the fixed extension ensemble and recorded. To obtain force values, generalized momenta are assigned to the free coordinates based upon the assumption of equipartition. These can be translated into generalized velocities with the metric tensor.}
\label{fig:schematic}
\end{figure}

Our polymer model (Fig.~\ref{fig:schematic}) consists of a chain of $N-1$ segments of equal length $a$, defined by the $N$ end points of each segment. The starting point of the $i$th segment (and ending point of the $i-1$th) will be defined as $\mathbf{r}_{i}$. This gives us a total of $3N$ Cartesian coordinates. $N-1$ inextensibility constraints ensure each displacement vector $\mathbf{d}_i$, defined as $\mathbf{r}_{i} - \mathbf{r}_{i-1} $, remains at a fixed length $a$. An additional constraint on the end-to-end extension leaves $2N$ free parameters. The potential $U$ will be defined in terms of the bending angle (the change in the tangent vector) between consecutive segments.
\begin{equation}
 U = \sum_{i=1}^{N-1} \frac{k_B T}{2} \left ( \frac{L_p}{a} \right ) \theta_i^2 \mbox{ where }
 \theta_i = \arccos(\mathbf{d}_i\cdot \mathbf{d}_{i+1}),
\end{equation}
where $L_p$ is the persistence length, equivalent to the macroscopic bending stiffness.
Ultimately, we seek the probability density function (PDF) of the force ($f$), which depends on the generalized coordinates ($\mathbf{q}$) and momenta ($\mathbf{p}$) according to
\begin{equation}
   f=\lambda(\mathbf{p},\mathbf{q}), 
\end{equation}
where the set of independent coordinates themselves follow a density function ($\rho$) according to a Boltzmann distribution
\begin{equation}
    \rho(\mathbf{p},\mathbf{q}) = \frac{e^{-\beta\mathcal{H}(\mathbf{p},\mathbf{q})}}{Z}
    \mbox{ where }
    Z = \int e^{-\beta\mathcal{H}(\mathbf{p},\mathbf{q})} d^{2N}q d^{2N}p,
\end{equation}
where $\mathcal{H}$ is the Hamiltonian, and $\beta$ is the inverse temperature.
The force PDF is then given by
\begin{equation}
    p(f)=\int \delta(f-\lambda(\mathbf{p},\mathbf{q}))\rho(\mathbf{p},\mathbf{q})d^{2N}q d^{2N}p.
\end{equation}
Since $\lambda$ is not bijective, the inverse function does not exist, and the integral for $p(f)$ cannot be calculated analytically. Therefore, to calculate $p(f)$, we must use numerical means. 

Calculating the constraint force $\lambda$ that maintains the chain at fixed extension requires both position and velocity information for all of the unconstrained coordinates. Sampling the velocity or momentum distribution for this system poses an additional challenge. Traditional Monte Carlo moves for isometric ensembles, such as crankshaft or backrub  moves~\cite{frank1985torsional,kumar1988off,betancourt2011optimization}, do not represent a full set of generalized coordinates because they are comprised of overlapping sets of angles. This requires us to find a new set of fully independent generalized coordinates, if they are to have well-defined partial derivatives and conjugate momenta. 

Once we have found such a set of coordinates, we employ a two-stage hybrid method schematized in Fig.~\ref{fig:schematic}(A): position information is sampled along the horizontal axis by a Monte Carlo process, and momentum information is sampled along the vertical axis by Gaussian sampling. Beginning from some initial state with the specified extension, one of the generalized coordinates is chosen at random and perturbed. The change in bending energy is computed, but an additional term is required to account for the relative size of the momentum space (Fig.~\ref{fig:schematic}(A)). This term gives a weighting factor that is included when we evaluate the Metropolis criterion. The result is the same as the average value we would get from including the kinetic energy in our Monte Carlo step, as this energy generally can depend on both position and momentum coordinates.

%This comes from the determinant of the metric tensor. Rather than computing all $2N$ columns of the Jacobian and metric for each new conformation, we can speed up this process by computing the determinant of the smaller $N\times N$ constraint metric~\cite{fixman1974classical}.

Next for momentum information, we use the Gaussian sampling method\cite{czapla2006sequence,agrawal2008geometry} which is more efficient than a Monte Carlo method. Coupling between momentum coordinates due to the length constraints, however, prohibits direct application of this method. Thus, we employ modal coordinates which are a useful tool for applying equipartition\cite{jain_equipartition_2012}. Mathematically, the kinetic energy ($KE$) of the system is given by a quadratic form
\begin{equation}
    KE=\frac{1}{2}\mathbf{p}^\text{T}\mathbf{M}^{-1}(\mathbf{q})\mathbf{p},
\end{equation}
where $\mathbf{M}^{-1}$ is the inverse of the mass matrix $\mathbf{M}$, and $\mathbf{q}$ and $\mathbf{p}$ are column vectors of generalized coordinates and momenta, respectively. In tensor form, $\mathbf{M}$ is defined in terms of the Cartesian coordinates $\mathbf{r}_k$ of each point mass $m_k$ as 
\begin{equation}
M_{ij} = \sum_{k=1}^N m_k \frac{\partial \mathbf{r}_k}{\partial q^i}\cdot\frac{\partial \mathbf{r}_k}{\partial q^j}, 
\label{eq:metric}
\end{equation}
which is simply referred to as the metric tensor. $\mathbf{M}^{-1}$ is not diagonal in general, but it is symmetric and positive definite. Thus it can be factored using a Cholesky decomposition into a triangular matrix $\bm{\mu}$ and its transpose. As a result, $KE$ can be brought to a diagonal form with respect to modal coordinates $\bm{\nu}$. 
\begin{equation}
    KE=\frac{1}{2}\mathbf{p}^\text{T}\bm{\mu}^\text{T}\bm{\mu}\mathbf{p}\equiv\frac{1}{2}\bm{\nu}^\text{T}\bm{\nu}.
\end{equation}
Since $KE$ obeys the Boltzmann distribution, components of $\bm{\nu}$ can be chosen from a normal distribution with a width given by the equipartition theorem.
\begin{equation}
    \langle \nu_i \nu_j\rangle = k_BT \delta_{ij}
\end{equation}
These are then converted into generalized momenta by back-substitution into the factored metric $\bm{\mu}$.

\subsection{Hierarchical Coordinates}

We consider a set of $2N$ generalized coordinates in a hierarchical fashion. At the highest level, three coordinates will describe large-scale movements of the chain. Remaining coordinates will only describe motions within one or the other half of the chain- they can be defined recursively with a set of fixed-extension coordinates being defined for each subchain as they were for the global system.

For simplicity, we assume the end-to-end vector is oriented along the $z$-axis. Defining the extension of the entire chain as $L_1$, we decompose it into two segments of length $L_{10}$ and $L_{11}$ (Fig.~\ref{fig:coordinates}). The coordinate $\phi_1$ defines the azimuthal angle of the chain about the axis connecting its endpoints. The angles $\theta_{10}$ and $\theta_{11}$ are derived from the coordinates $L_1, L_{10}$ and $L_{11}$, and represent the polar angle of the segments described by $L_{10}$ and $L_{11}$ relative to the axis defined by $L_1$. These are expressed as
\begin{equation}
\theta_{10} = \arccos\left( \frac{L_1^2+L_{10}^2-L_{11}^2}{2L_1 L_{10}}\right)
\mbox{ , }
\theta_{11} = \arccos\left( \frac{L_1^2-L_{10}^2+L_{11}^2}{2L_1 L_{11}}\right)
\end{equation}
relative to the axis of the entire chain, in a plane determined by the angle $\phi_1$. This azimuthal angle $\phi_1$, along with $L_{10}$ and $L_{11}$, comprise the three coordinates at this level. If $L_{10}$ and $L_{11}$ represent single links, then those extensions are held fixed. If they represent multiple links, then they can contract or extend and the $\theta$ angles will change accordingly.

\begin{figure}[h]
\includegraphics[width=11.25cm]{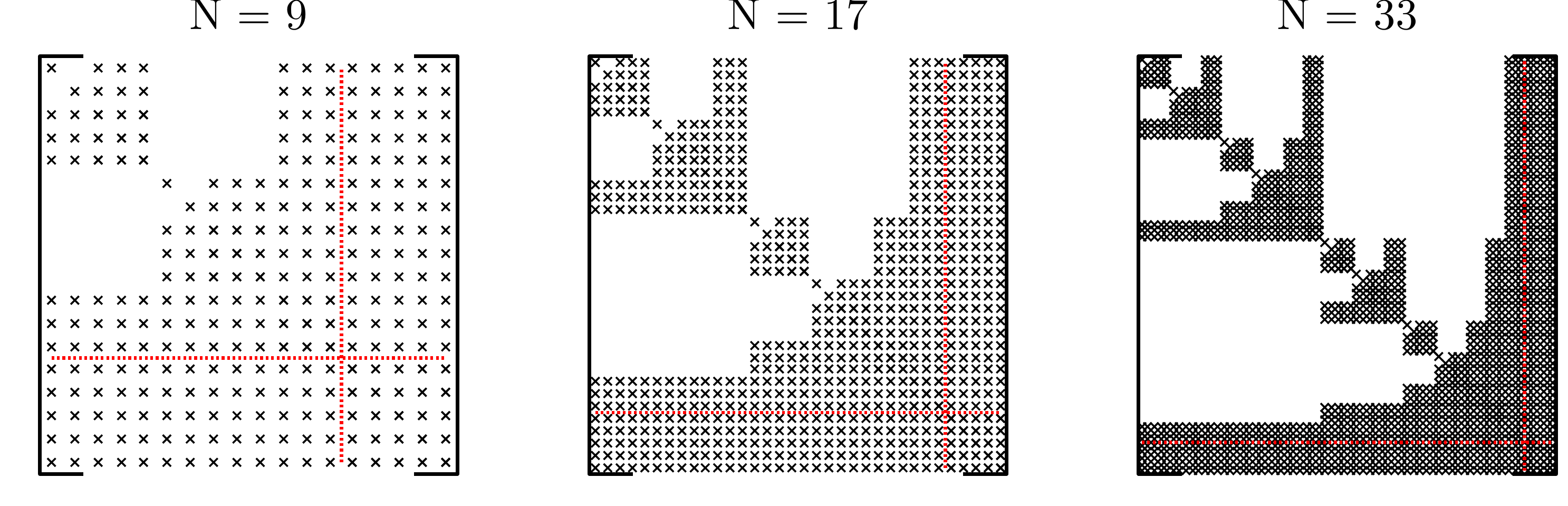}
\caption{Schematic of non-zero matrix elements for different values of $N$. Blank spaces represent values which are always zero, $\times$'s represent values which may be non-zero. The red dashed line separates the internal and global coordinates. The pattern of non-zeros at $N=2^{l}+1$ is repeated twice within the pattern for $N=2^{l+1}+1$, creating a fractal structure. }
\label{fig:matrix_schematic}
\end{figure}

At the global level, we will have five additional coordinates- three translations and two rotations of the end-to-end axis. The advantage of these hierarchical coordinates is that the resulting metric tensor will be sparse. While the size of the tensor will scale as $N^2$, the number of non-zero entries will scale as $N\log N$ (Fig.~\ref{fig:matrix_schematic}). This will greatly expedite computing the matrix and its derivatives.

\subsubsection{Crankshaft Rotation Moves}

A crankshaft rotation move alters one of the azimuthal angles $\phi$. On a set of points, it will be defined by rotating the interior points about the axis connecting the end points. An example is found in Fig.~\ref{fig:coordinates}(A).
This will preserve all the interior distances, and the overall end-to-end vector for the subchain. Crankshaft moves may serve as part of a complete set of generalized coordinates, provided the intervals they span do not partially overlap. Crankshaft angles of disjoint subchains, or a subchain that is entirely contained within another, may be altered independently, but angles for partially overlapping subchains may not.

\subsubsection{Expansion Moves}

\begin{figure}[th!]
\includegraphics[width=15cm]{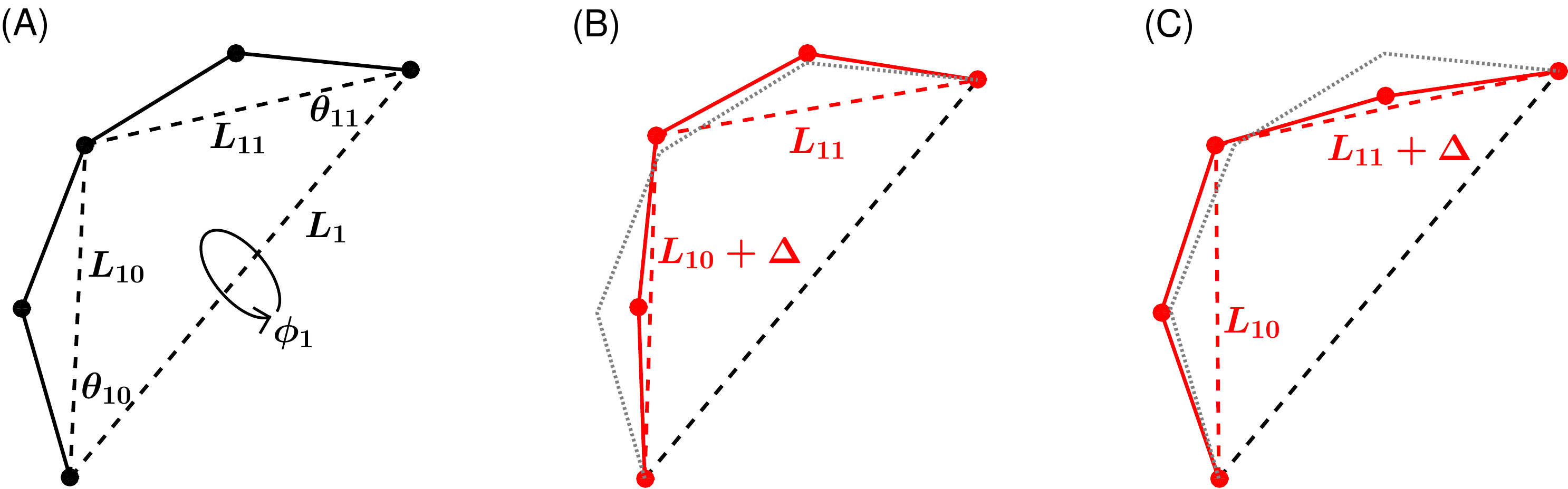}
\caption{Generalized coordinates at one level. In (A), the crankshaft rotation preserves all internal end-to-end distances. In (B), the extension $L_1$ is preserved while $L_{10}$ is altered. In (C), $L_1$ remains fixed while the extension $L_{11}$ is changed. The same set of coordinates will exist within the subchains $L_{10}$ and $L_{11}$.}
\label{fig:coordinates}
\end{figure}

Expansion moves will come in two varieties. Defining the midpoint of a set of vertices $\mathbf{r}_i\dots \mathbf{r}_{i+l}$ as $\mathbf{r}_{i+k}$ where $k$ is halfway to $l$, rounded up ($k = \lceil l/2\rceil$), one move will expand or contract the points $\mathbf{r}_i\dots \mathbf{r}_{i+k}$ by changing the angle at their midpoint $\mathbf{r}_{i+j}$ where $j = \lceil k/2\rceil$ while simultaneously rotating the points $\mathbf{r}_{i+k+1}\dots \mathbf{r}_{i+l-1}$ about the point $\mathbf{r}_{i+l}$ to preserve the interior distances $|\mathbf{r}_i - \mathbf{r}_{i+l} |$ and $|\mathbf{r}_{i+l} - \mathbf{r}_{i+k} |$, as in Fig.~\ref{fig:coordinates} B. The other expansion move will preserve the distances  $|\mathbf{r}_i - \mathbf{r}_{i+l} |$ and $|\mathbf{r}_i - \mathbf{r}_{i+k} |$ while expanding or contracting the chain between $\mathbf{r}_{i+k}$ and $\mathbf{r}_{i+l}$, as in Fig.~\ref{fig:coordinates} C. These expansion moves are similar to other algorithms based on solving the inverse kinematic problem\cite{nilmeier2011assessing,bottaro2012subtle,zamuner2015efficient}, which work by applying a stochastic rotation step on one segment and applying a deterministic rotation step on another segment to close the chain.

We can then find the number $m$ of generalized internal coordinates for a chain of $N-1$ links and $N$ points, using a recursive formula  
\begin{equation}
m(N) = \begin{cases}
0 & N < 3 \\
1 & N = 3 \\
2 + m(3) & N = 4 \\
3 + m(\lceil N/2 \rceil) + m(\lfloor N/2 \rfloor) & N > 4
\end{cases}
\end{equation}
This can easily be shown to yield $m(N) = 2N-5$ for all $ N \ge 3$. Adding in the five global coordinates gives a full set of $2N$ generalized coordinates.

\subsection{Monte Carlo Step}
The conformational space at fixed extension is explored by randomly selecting one coordinate and perturbing it by a normally-distributed random value. This can be done in phase-space, perturbing either a position or momentum coordinate. 
However, to improve performance, we can consider only perturbations in position space. Integrating over momentum coordinates weighted by kinetic energy, leaves us with the square root of the determinant of the covariant metric tensor ($M^{1/2}$). The Metropolis criterion of acceptance probability ($P_{q\rightarrow q'}$)  will then take the form
%The Metropolis criterion of acceptance probability ($P$) for a position perturbation ($q\rightarrow q'$) will take the form
%\begin{equation}
%P_{q\rightarrow q'} = \begin{cases}
%1  & U(q') + \frac{1}{2}p_i M^{ij}(q') p_j \le U(q) + \frac{1}{2}p_i M^{ij}(q) p_j %\\
%\exp(-\beta(\Delta U + \Delta T))  & U(q') + \frac{1}{2}p_i M^{ij}(q') p_j > U(q) %+ \frac{1}{2}p_i M^{ij}(q) p_j
%\end{cases}
%\end{equation} 
\begin{equation}
P_{q\rightarrow q'} = \begin{cases}
1  & U(q') -\frac{1}{2}\log{M(q')} \le U(q) - \frac{1}{2}\log{M(q)} \\
\exp(-\beta \Delta U) \sqrt{\frac{M'}{M}} & U(q') -\frac{1}{2}\log{M(q')} > U(q) - \frac{1}{2}\log{M(q)}
\end{cases}
\end{equation}
%. The change in kinetic energy $\Delta T = \frac{1}{2}p_i M^{ij}(q') p_j - \frac{1}{2}p_i M^{ij}(q) p_j $ comes from the way the contravariant form of the metric tensor $M^{ij}$ changes with the position coordinates. The derivation of this tensor from the generalized coordinates is discussed below.
%For a momentum perturbation ($p\rightarrow p'$), the potential energy remains constant and we have
%begin{equation}
%P_{p\rightarrow p'} = \begin{cases}
%1  & \frac{1}{2}p'_i M^{ij} p'_j \le \frac{1}{2}p_i M^{ij} p_j\\
%\exp(- \beta \Delta T)  & \frac{1}{2}p'_i M^{ij} p'_j > \frac{1}{2}p_i M^{ij} p_j 
%\end{cases}
%\end{equation}
%To improve performance, we can consider only perturbations in position space. Integrating over momentum coordinates, weighted by kinetic energy, leaves us with a factor $M^{1/2}$ equal to the square root of the determinant of the covariant metric tensor.
where the change in total bending energy $\Delta U=U(q')-U(q)$ is calculated at all pivot points for the perturbation. Rather than computing the determinant $M$ of the large matrix of free coordinates, its inverse can be obtained more efficiently using a smaller tridiagonal matrix $\mathbf{H}$ in the constrained coordinates\cite{fixman1974classical}. For an inextensible polymer, these constraints correspond to the fixed length of each segment, $\|\mathbf{r}_i-\mathbf{r}_{i+1} \| = a_i$, 
which results in a tridiagonal matrix
\begin{equation}
    H_{ij} = \sum_{l=1}^{N-1} \mathbf{\nabla_l} a_i \cdot \nabla_l a_j .
\end{equation}
Applying the method to a chain with fixed end-to-end distance introduces one additional constraint, $\|\mathbf{r}_N-\mathbf{r}_{1} \| = a_N$. This will result in a matrix $\mathbf{H}'$ which is tridiagonal with the exception of two entries in the $\{1,N\}$ and $\{N,1\}$ matrix elements.

\subsection{Computing Forces}

The constraint force with respect to the end-to-end distance can be found from the Lagrangian for the polymer model. The Lagrangian $\mathcal{L}$ in the $2N+1$ dimensional space of free coordinates $q^i$ and the constrained extension $q^\xi$ is given by
\begin{equation}
\mathcal{L} = \frac{1}{2}\dot q^i \mathcal{M}_{ij}q^j - U(q^i) - \lambda (q^\xi-r)
\end{equation}
where $\lambda$ is an undetermined multiplier corresponding to the constraint force and $r$ is the constrained value of the end-to-end distance.
%We can define a metric tensor for the generalized coordinates by taking an inner product over the derivative of all Cartesian coordinates with respect to the generalized coordinates $q^i$, where the position of the point mass $m_k$ is given by the vector $\mathbf{r}_k$. Here the index $i$ ranges over both the $2N$ free coordinates, and the constrained coordinate denoted by $q^{\xi}$.
%\begin{equation}
%\mathcal{M}_{ij} = \sum_{k=1}^N m_k \frac{\partial \mathbf{r}_k}{\partial q^i}\cdot\frac{\partial \mathbf{r}_k}{\partial q^j}
%\end{equation}
%his allows us to define a kinetic energy in terms of the generalized coordinates, and from this a Lagrangian for the $2N+1$ dimensional system. 
%where $\lambda$ is an undetermined multiplier corresponding to the constraint force and $q^\xi_0$ is the constrained value of the end-to-end distance. 
The metric ($\mathcal{M}_{ij}$) here is in the larger, $2N+1$ dimensional space. Represented in block form, this is equivalent to 
\begin{equation}
\mathcal{M} = \left[
\begin{array}{c c}
 A & \mathbf{B}^T \\
 \mathbf{B} & \mathbf{M}
\end{array}
\right]
\end{equation}
where 
\begin{equation}
    B_{i} = \sum_{k=1}^N m_k \frac{\partial \mathbf{r}_k}{\partial q^i}\cdot\frac{\partial \mathbf{r}_k}{\partial q^\xi}  \mbox{ and }
    A = \sum_{k=1}^N m_k \frac{\partial \mathbf{r}_k}{\partial q^\xi}\cdot\frac{\partial \mathbf{r}_k}{\partial q^\xi} 
\end{equation}

Making use of the symmetry of the metric tensor, the equation of motion for the  coordinate $q^\xi$ is
\begin{equation}
\ddot q^i  \mathcal{M}_{i\xi}+\dot q^i \dot q^j\frac{\partial \mathcal{M}_{i\xi}}{\partial q^j} = \frac{1}{2} \dot q^i \dot q^j \frac{\partial  \mathcal{M}_{ij}}{\partial q^\xi}
-\frac{\partial U}{\partial q^\xi} -\lambda
\end{equation}
Since $q^\xi$ is constrained, $\dot q^\xi$ is zero.
The remaining generalized velocities can be initialized using the modal velocity scheme outlined above.  Using the terms defined in our block representation, this can be rewritten as
\begin{equation}
\ddot q^i B_{i}+\dot q^i \dot q^j\frac{\partial B_{i}}{\partial q^j} = \frac{1}{2} \dot q^i \dot q^j \frac{\partial  M_{ij}}{\partial q^\xi}
-\frac{\partial U}{\partial q^\xi} -\lambda
\end{equation}
Obtaining $\lambda$ will also require the generalized accelerations of the unconstrained coordinates, which can be found from the remaining $2N$ equations of motion
\begin{equation}
    \ddot q^i M_{ik}+\dot q^i \dot q^j\frac{\partial M_{ik}}{\partial q^j} = \frac{1}{2} \dot q^i \dot q^j \frac{\partial M_{ij}}{\partial q^k}
-\frac{\partial U}{\partial q^k} 
\end{equation}
In terms of the contravariant form of the metric tensor $M^{ij}$ defined by $M_{ij}M^{jk}=\delta_i^k$, we can express the constraint force as
\begin{equation}\label{eq:constraint}
\lambda  =  \underbrace{-M_{ik}\dot q^i \dot q^j\frac{\partial B^{k}}{\partial q^j} + \frac{1}{2} \dot q^i \dot q^j \left( \frac{\partial}{\partial q^\xi} -B^k\frac{\partial }{\partial q^k} \right) M_{ij}}_{\mbox{Inertial}} - \underbrace{\left( \frac{\partial}{\partial q^\xi} -B^k\frac{\partial }{\partial q^k} \right) U}_{\mbox{Elastic}}
\end{equation}
where the vector $B_i$ is converted from covariant to contravariant using the metric $M^{ij}$ on the unconstrained subspace. The first and second terms, which depend on the kinetic energy of the polymer chain, are categorized as the inertial force (or entropic force in an average sense), whereas the third term proportional to the bending energy with no velocity dependence is categorized as the elastic force.

%The partial derivative $\frac{\partial}{\partial q^\xi}$ represents change of the potential energy or metric along a direction that preserves all other coordinates, but we must bear in mind that those coordinates are arbitrary and typically not orthogonal to the constrained coordinate $q^\xi$ or each other. The combined term $\frac{\partial}{\partial q^\xi} -B^k\frac{\partial }{\partial q^k}$ represents a derivative along the direction normal to surfaces of constant $\xi$, independent of our choice of coordinates. 

\section{Results}
\subsection{Constraint Forces vs. Generalized Force}
\begin{figure}[th!]
\includegraphics[width=8.25cm]{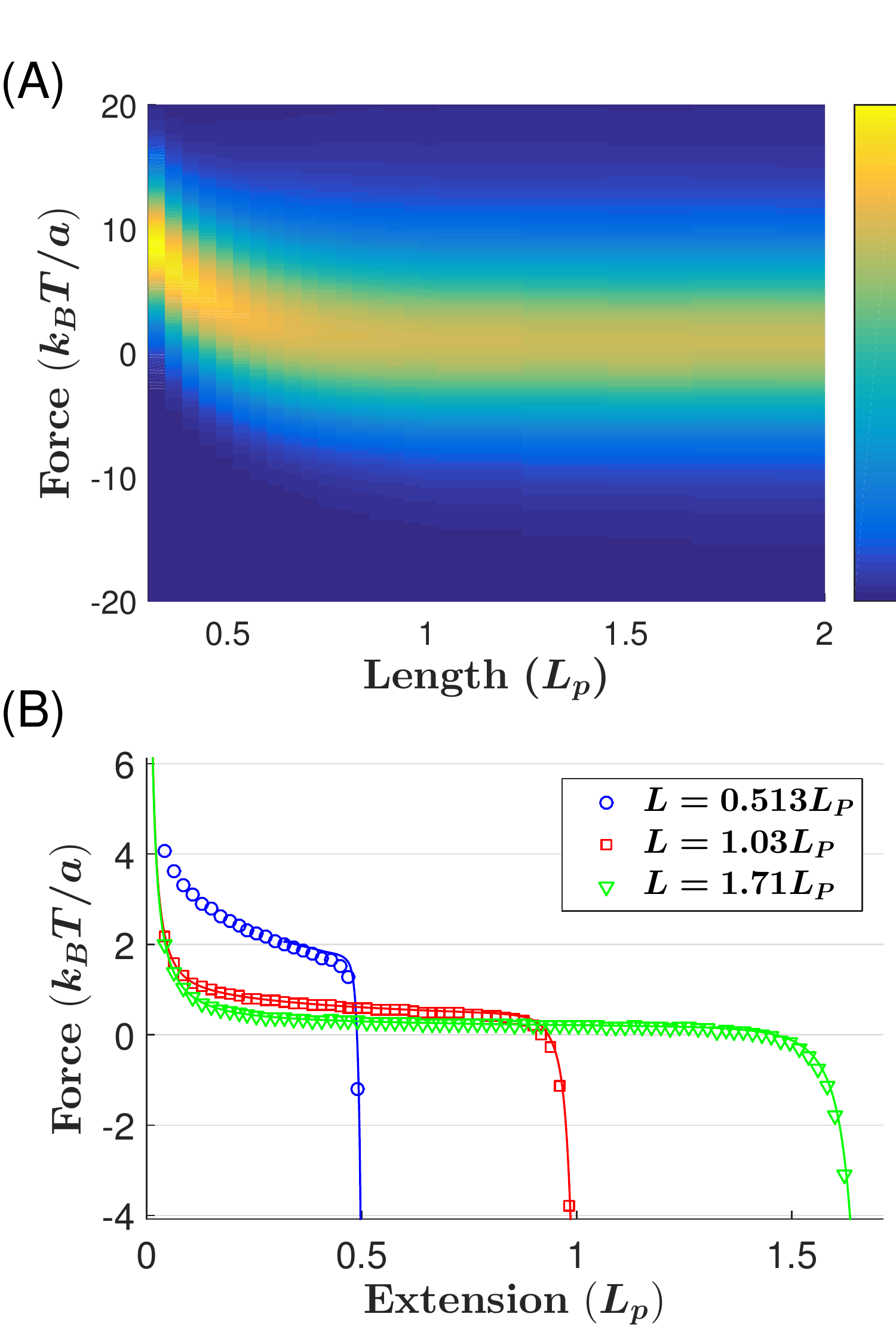}
\caption{Constraint forces. (A) Distribution of forces as a function of length. The PDF values are plotted as a heat map using the colormap shown on the right. Values are at a fixed extension of 0.068 $L_p$. As longer chains are considered, the average force shifts towards zero, and the distribution grows broader. (B) The generalized force ($\overline f$) obtained from partition function (Eq.~\ref{eq:meanforce}), alongside the mean constraint force ($\langle \lambda \rangle$) obtained from our phase space sampling method. The two methods show good agreement over a range of extensions, though the computation of the end-to-end distribution becomes unstable for short chains at short extension.}
\label{fig:fdist}
\end{figure}

Using the phase-space sampling method, we calculated instantaneous constraint forces $\lambda$ exerted by a semiflexible chain held at constant extension $r$. The full distribution of $\lambda$ as a function of contour length is plotted as a heat map in Fig.~\ref{fig:fdist}A. As expected, the mean force ($\langle \lambda \rangle$) decreases with increasing chain length. To check the validity of this result, we compared $\langle \lambda \rangle$ to the generalized force $\overline f$ from thermodynamics. The generalized force $\overline f$ conjugate to $r$ can be derived from the free energy of the looped macrostate ($A(r)$) according to
\begin{equation}
\overline{f}=\frac{\partial A(r)}{\partial r}=-k_BT\frac{\partial \log{P(r)}}{\partial r},
\label{eq:meanforce}
\end{equation}
where $P(r)$ is the PDF of $r$. In Fig.~\ref{fig:fdist}B, the forces calculated as a function of extension for three different contour lengths are shown. The mean constraint force ($\langle \lambda \rangle$) is shown in hollow symbols, and the generalized force ($\overline f$) is shown in solid lines. The generalized force was calculated using a semi-analytical expression for $P(r)$ derived by Mehraeen et al.~\cite{mehraeen2008end}. As shown, the two methods produce good agreement across different extensions and contour lengths. This confirms that in terms of average force, our phase-space sampling method is consistent with the prediction of statistical mechanics.

%
%\begin{figure}[h]
%\includegraphics[width=8.25cm]{fig_rdf_force.pdf}
%\caption{(A) Distribution of end-to-end distances ($P(r)$) obtained from the worm-like chain model. For chains less than one persistence length, this is sharply peaked near the length of the chain. As the contour length increases, the distribution spreads out. (B) Mean force ($\bar f$) obtained from partition function (Eq.~\ref{eq:meanforce}), alongside that ($\langle f \rangle$) obtained from our phase space sampling method. The two methods show good agreement over a range of extensions, though the computation of the end-to-end distribution becomes unstable for short chains at short extension.}
%\label{fig:rdf_force}
%\end{figure}

\subsection{Kinetic and Potential Contributions to the Mean Force}
The mean force $\langle \lambda \rangle$ from the semiflexible loop is positive for short extensions, which indicates that the ends of the loop must be pulled inward to keep the end-to-end distance constant. This outward direction of the force is intuitively predictable based on the force required to maintain a macroscopic elastic rod in a deformed state. More quantitatively, in the absence of thermal fluctuations, the minimum energy conformation of an elastic rod with a short fixed end-to-end distance is a teardrop which needs to be held with a tensile force\cite{allemand2006loops}. 

\begin{figure}[h]
\includegraphics[width=17.0cm]{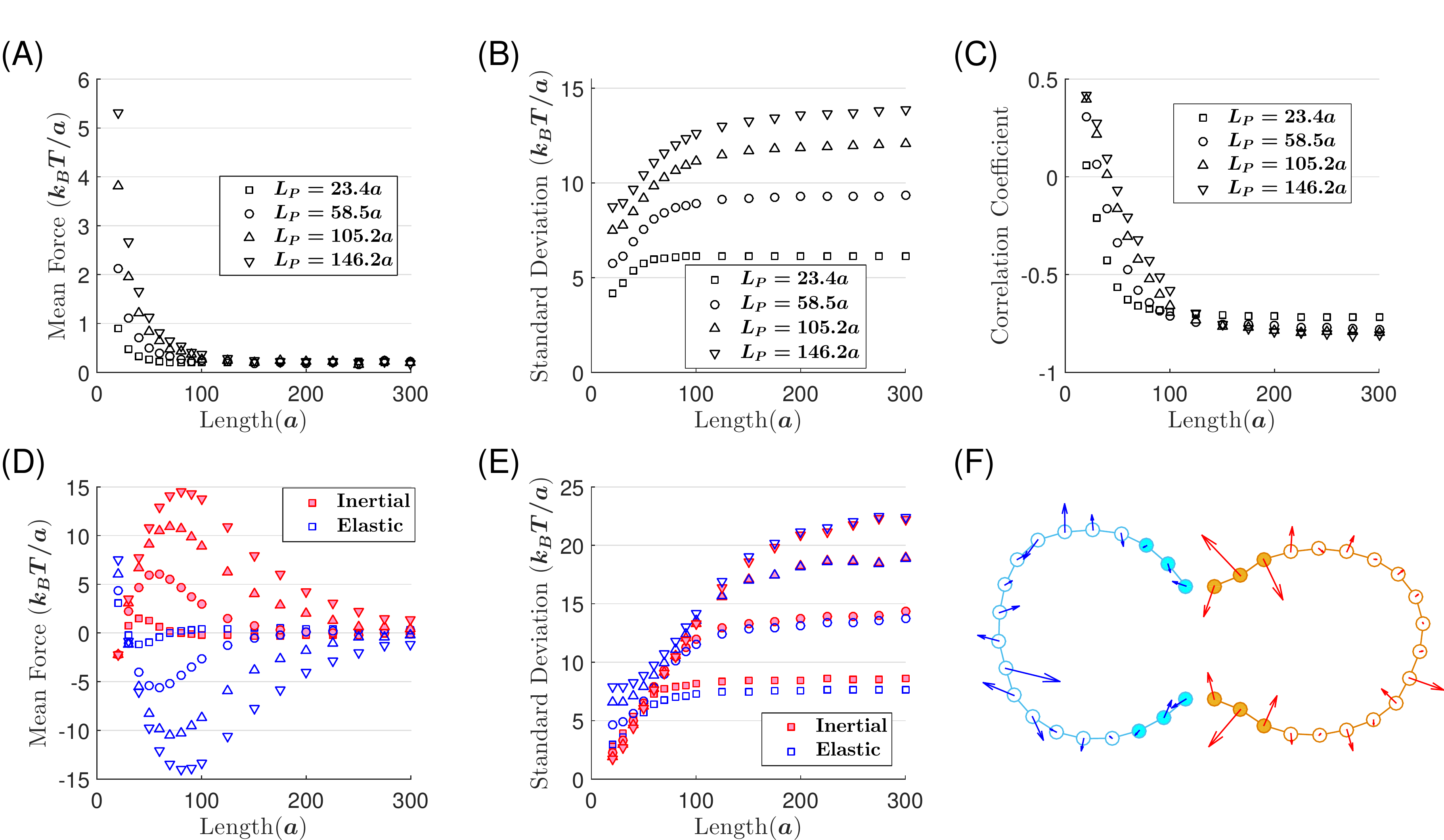}
\caption{Total constraint force vs. its elastic and inertial components. (A) Average force decreases with contour length, and increases with persistence length. (B) Standard deviation increases with persistence length and contour length, plateauing around $L_p$.
(C) Correlation coefficient decreases with length, with inertial and elastic components moving from correlated to anticorrelated.
(D) Decomposition of force into potential and kinetic components versus contour length, at a constant extension of $10 a$. As contour length increases, the contribution from the bending potential asymptotes to zero. The non-zero force at large contour lengths is a result of entropic contributions only. Also of note is the fact that the magnitude of both individual components generally increases with persistence length. (E) Deviation of force components follows the same trend as their sum, but reaches a larger value.  (F) Two example conformations, taken from a two-dimensional ensemble with  length $20a$ and end-to-end separation $4 a$. Arrows indicate the instantaneous acceleration arising from the bending potential.
In the conformation on the left (light blue) the first and last point are moving apart, and a linker between the two ends will be subject to a stretching force. In the conformation on the right (orange) these points are moving closer together and the linker between the two ends will be compressed. While these forces are a product of all the coordinates, the separation into stretching or compressing forces is strongly correlated with the convexity or concavity in the first three mass points from the end.}
\label{fig:PK_Decomposition}
\end{figure}

However, our simulation reveals that this macroscopic-level understanding does not always apply to a thermally-excitable semiflexible loop. The mean force can be dissected into an elastic force that arises from the internal energy stored in the deformed chain and an entropic force that arises from the inertia of moving mass points\cite{waters2015calculation}. The mean elastic force is mostly negative, thus compressive rather than tensile (blue hollow symbols, Fig.~\ref{fig:PK_Decomposition}D). This negative force is compensated by a slightly larger positive entropic force (red filled symbols, Fig.~\ref{fig:PK_Decomposition}D) to yield a net positive mean force. Shown in Fig.~\ref{fig:PK_Decomposition}F are example conformations that produce positive (left) and negative (right) elastic force. The conformations with negative elastic forces typically exhibit inflection points in the contour near the ends such that the end segments bent outward exert a compressive elastic force along the end-to-end vector. Compressive elastic forces between the ends of an elastic chain are not intuitive, but can be demonstrated even at the macroscopic level \cite{bosi2015development}.

\subsection{Effects of Stiffness and Length}
We investigated how the force profile changes with two chain parameters, stiffness ($L_p$) and contour length ($L$) while keeping the end-to-end distance constant. As the length increases, the mean force decreases, but does not reach zero even at large contour lengths (Fig.~\ref{fig:PK_Decomposition}A). In contrast to the total mean force, the elastic and entropic forces change nonmonotonically with length. At very short lengths (less than 30 monomer lengths), the entropic force is compressive, and the elastic force is tensile. But beyond this length, they reverse signs and grow in magnitude with increasing length. In this regime, both the compressive elastic force and tensile entropic force increase in magnitude with increasing stiffness. 

As the contour length goes up, the amplitude of fluctuations rises and plateaus on the scale of one persistence length (Fig.~\ref{fig:PK_Decomposition}B). This behavior is similarly followed by both elastic and inertial force fluctuations (Fig.~\ref{fig:PK_Decomposition}E). This implies that large force values occur more frequently, even as the average force goes down. 
We also calculated the correlation coefficient between the elastic and inertial forces (covariance) as a function of length and stiffness (Fig.~\ref{fig:PK_Decomposition}C). We see a crossover from positive to negative correlation around 50 monomer lengths. The negative correlation increases and plateaus on a similar scale to the fluctuations. This negative correlation implies that the fluctuation of the sum of inertial and elastic components (Fig.~\ref{fig:PK_Decomposition}B) is less than the fluctuation of such components considered individually (Fig.~\ref{fig:PK_Decomposition}E).

Using the same method, we also explored the effect of chain extension $r$ on the force distribution at four different contour lengths. We chose four different contour lengths, ranging from $0.34 L_p$ up to $1.7 L_p$. Extension-to-contour ratio, which is between 0 and 1, tells us whether the chain is loop-like or rod-like. In the short extension (loop-like) regime, the average force strongly favors larger extensions as result of entropic effects.
%These fall off sharply, though, 
The force fluctuation also decreases as shown by the narrowing of contour lines. In the intermediate extension regime, the average force varies slowly, and the fluctuation of the force is also stabilized with contour lines forming a bottleneck-like pattern. At extensions near the contour length (rod-like), the average force takes on the opposite sign because completely straight conformations are unfavorable due to entropy. In this rod-like regime, the force fluctuation diverges rapidly.  

\begin{figure}[h]
\includegraphics[width=11.25cm]{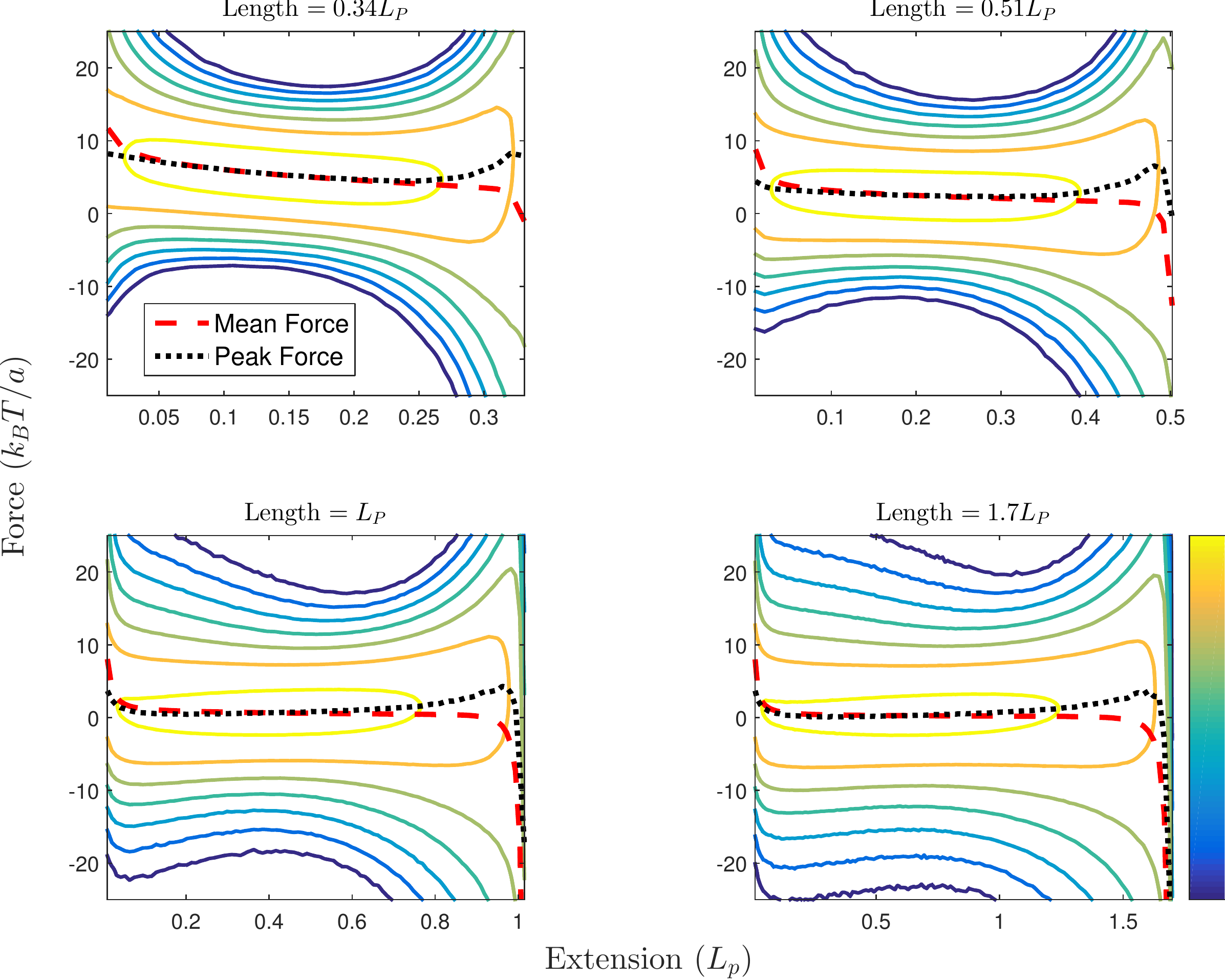}
\caption{Contour plots of force distribution vs extension for four different chain lengths. Each contour line is colored according to its corresponding PDF amplitude. All figures represent data sampled for chains with a persistence length $L_p = 23.4 a$. The peak value changes rapidly in regions of low and high extension, but slowly in intermediate regions. These intermediate regions also correspond to the narrowest distribution of forces. }
\label{fig:force_extension}
\end{figure}
%The mean force corresponds to the derivative of the free energy obtained from the partition function. This partition function corresponds to the end-to-end distribution for a continuous chain, and so representing the same contour length (expressed as a number of persistence lengths) with smaller or larger numbers of discrete segments will yield the same result, after rescaling to account for the different monomer length between the discretized systems. Representing a chain with a smaller or larger number of monomers per persistence length simply requires scaling the elastic bending energy at each junction down or up.

%While the average force goes down as persistence length increases, the deviation of the forces goes up. Fitting a power law to the fluctuation for large contour lengths reveals that they are related as $\sigma \propto L_p^{0.44}$, close to the square root law for fluctuations in the toy model. The reason for this increase is that a longer persistence length means more monomers (and more degrees of freedom) for the same length of chain. Each degree of freedom has approximately the same thermal energy associated with it, and so chains of longer persistence length have more energy and larger fluctuations.

%\subsection{Distribution of Force Along Loop}

\subsection{Parameterizing Force Distribution}
\begin{figure}[h]
\includegraphics[width=8.5cm]{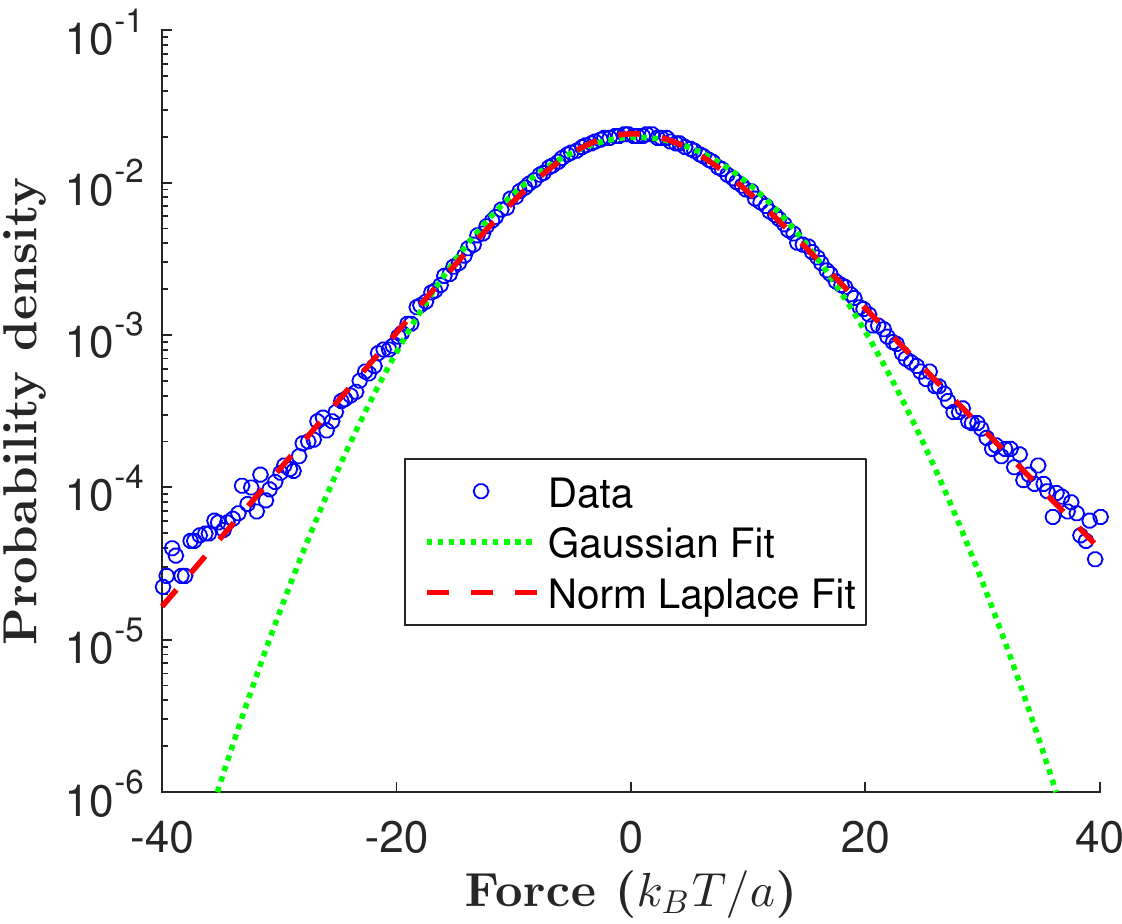}
\caption{Force distribution (blue circles) for $L = 60 a$ vs. Gaussian (green line) and Normal Laplacian (red dashed line) fit functions. This clearly displays the disagreement of the Gaussian fit far from the peak value.}
\label{fig:force_dist_length}
\end{figure}

% If we do nothing and display no parameters, this does not belong. JTW
The distribution of forces at different contour lengths, persistence lengths, and extensions falls off more gradually than a Gaussian distribution. Instead, we find that the distribution of forces is well-approximated by a two-sided exponential distribution. To capture the asymmetry of the distribution, as well as the smoothing about the peak, we employ a normal Laplace distribution~\cite{reed2004double}, corresponding to a convolution of an asymmetric Laplace distribution with a normal distribution. The cumulative distribution function (CDF) is given by
\begin{equation}
CDF(f) = \frac{\alpha\beta}{\alpha+\beta}\phi\left(\frac{f-\nu}{\tau}\right)\left[
R(\alpha\tau - (f-\nu)/\tau)-R(\beta\tau+(f-\nu)/\tau)
\right]
\end{equation}
where $\phi$ is a standard normal distribution, and $R$ is the Mills Ratio between the normal distribution and its
corresponding CDF.
\begin{equation}
R(x) = \frac{1-\int_{-\infty}^x \phi(x')dx'}{\phi(x)}
\end{equation}
This expression has four free parameters ($\alpha,\beta, \nu,\tau$), which can be fit using a maximum likelihood estimation technique. Initial values to begin the maximum likelihood search for these parameters can be obtained from the first four moments of the force distribution. The Gaussian and normal Laplace distribution fits are shown in Fig.~\ref{fig:force_dist_length}.

%\subsection{Free Energy and Mechanical Work}

\section{Discussion}

%summary of key findings
We investigated the force distribution in a semiflexible polymer held at a fixed extension. We used the Kratky-Porod wormlike chain to coarse-grain the system, and employed a novel phase-space sampling method to obtain thermally-equilibrated chain conformations satisfying the constraints. We showed that the force distribution produces a mean that matches the generalized force derived from thermodynamics. By analyzing the inertial and elastic contributions to the constraint force, we found that in loop-like geometry (short extension compared to contour length), the entropic force pulls the ends outward (tensile) while the elastic force pushes them inward (compressive), which is contrary to our intuition based on elastic deformation. Our approach allows access to the force distribution in greater detail than simply the mean value. The distribution is skewed and broad compared to a Gaussian distribution. Notably, at short extensions the mean of the distribution decreases with length whereas the width increases. The agreement between average mechanical force and force from free energy can be invoked to extend this method to situations where the partition function is not easily obtainable, such as DNA loops with sequence-dependent intrinsic shape and flexibility. The force distribution may prove useful for the prediction of looped-state lifetimes in cases where the loop can be destabilized by critical forces exceeding some threshold.

The average of our sampled mechanical force agrees well with the generalized force obtained from the partition function (Fig.~\ref{fig:fdist}) in spite of several differences between the ensembles under consideration. The ensemble for the partition function~\cite{mehraeen2008end} consists purely of spatial conformations of a continuously deformable chain without kinetic energy or velocity constraint on the end-to-end distance. In comparison, the ensemble for our constraint force includes both kinetic and potential energy information of a discrete chain constrained in both position and momentum coordinates of the end-to-end distance. The difference between conditional and constrained averages has been well studied in relation to constrained MD simulations\cite{sprik1998free,den1998calculation,schlitter2003free,den2013revisiting}, but does not seem to be noticeable for our coarse-grained model.

\begin{figure}[h]
\includegraphics[width=8.25cm]{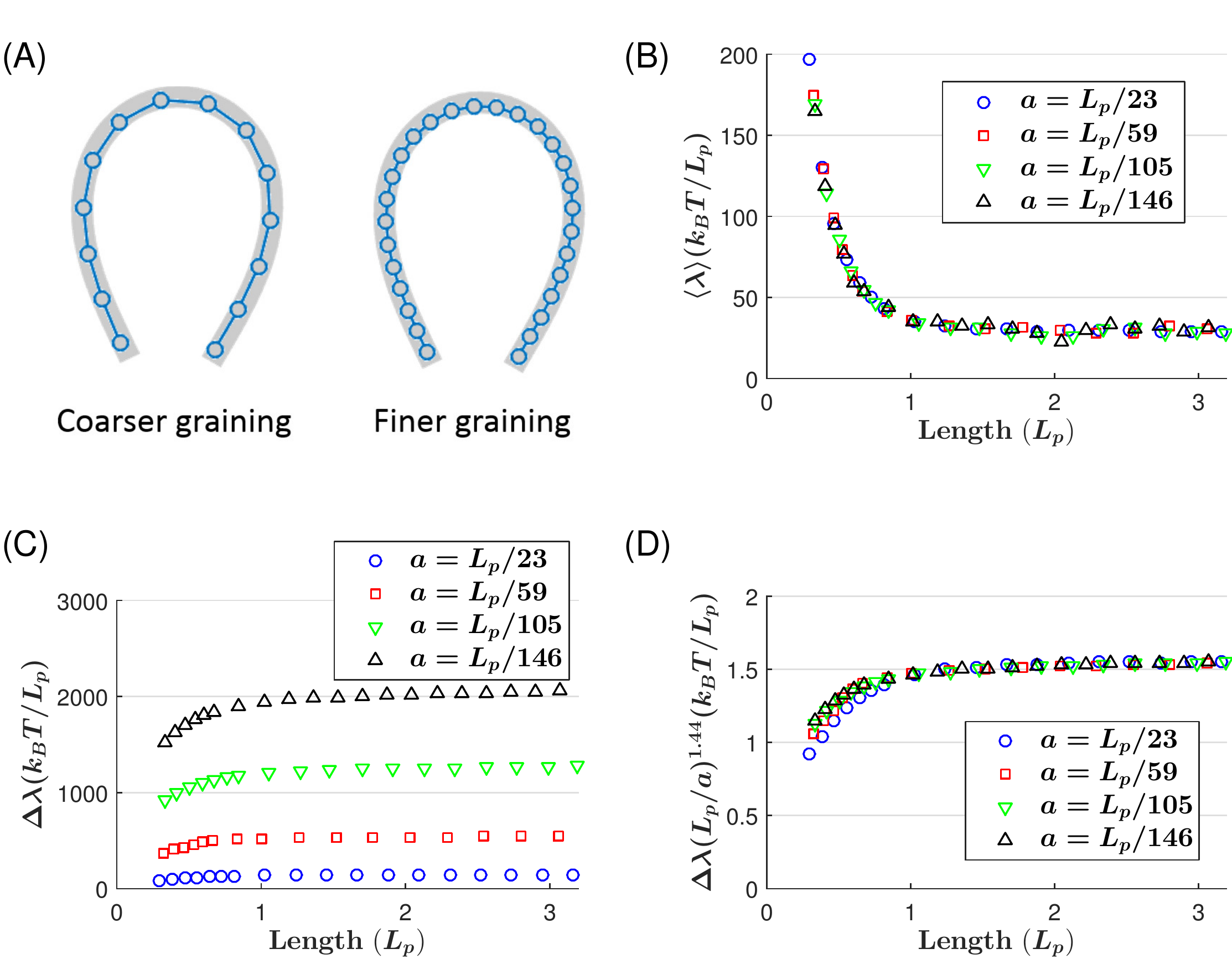}
\caption{The effect of coarse-graining on force distribution. (A) Diagram of coarse-graining process. A physical polymer with a given extension and persistence length can be represented with an arbitrary number of points in a ball-and-stick model. (B) Average Force at a fixed extension $r = 0.068 L_p$. As the chain increases in length, the average force decreases. The same trend, and same values, are predicted independent of number of points per persistence length. (C) Standard deviation of force at a fixed extension $r = 0.068 L_p$. The fluctuation increases as the chains grow longer. Additionally, the size of fluctuations increases as more points are used to represent the chain, as each degree of freedom corresponds to more thermal energy in the system. (D) Fluctuations scaled by $(L_p/a)^{1.44}$. This measured scaling factor accounts for the growth of force fluctuations with coarse graining.}
\label{fig:force_renorm}
\end{figure}

%\subsection{Scaling of Force Distribution}

As our simulation is based on a coarse-grained polymer with freedom on the level of coarse-graining, we can ask how our results depend on the choice of this free quantity. The granularity of coarse-graining is represented by the number $N$ of small length elements the polymer is divided into (Fig.~\ref{fig:force_renorm}A). By increasing $N$, both the monomer length $a$ and mass $m$ decrease. We computed the force distribution at different $N$, and found that the mean force $\langle \lambda \rangle$ does not change (Fig.~\ref{fig:force_renorm}B). In contrast, the standard deviation $\Delta \lambda$ increases with monomer number $N$ regardless of chain length (Fig.~\ref{fig:force_renorm}C). We found an approximate scaling law between $\Delta \lambda$ and $N$, where  $\Delta \lambda \sim N^{1.44}$. Dispersions normalized by $N^{1.44}$ roughly collapse to one curve (Fig.~\ref{fig:force_renorm}D). This result implies that the force fluctuation, unlike the mean force, increases with the degrees of freedom with no bound, similar to a Casimir-like force between two plates\cite{bartolo2002fluctuations}. However, due to the intrinsic microscopic length scale in a physical system, these degrees must be bounded at some level. Therefore, while the absolute value of fluctuation cannot be treated as universal, its behavior as a function of other chain parameters appears to be preserved across levels of coarse-graining (Fig.~\ref{fig:force_renorm}D). 

\begin{figure}[h]
\includegraphics[width=8.25cm]{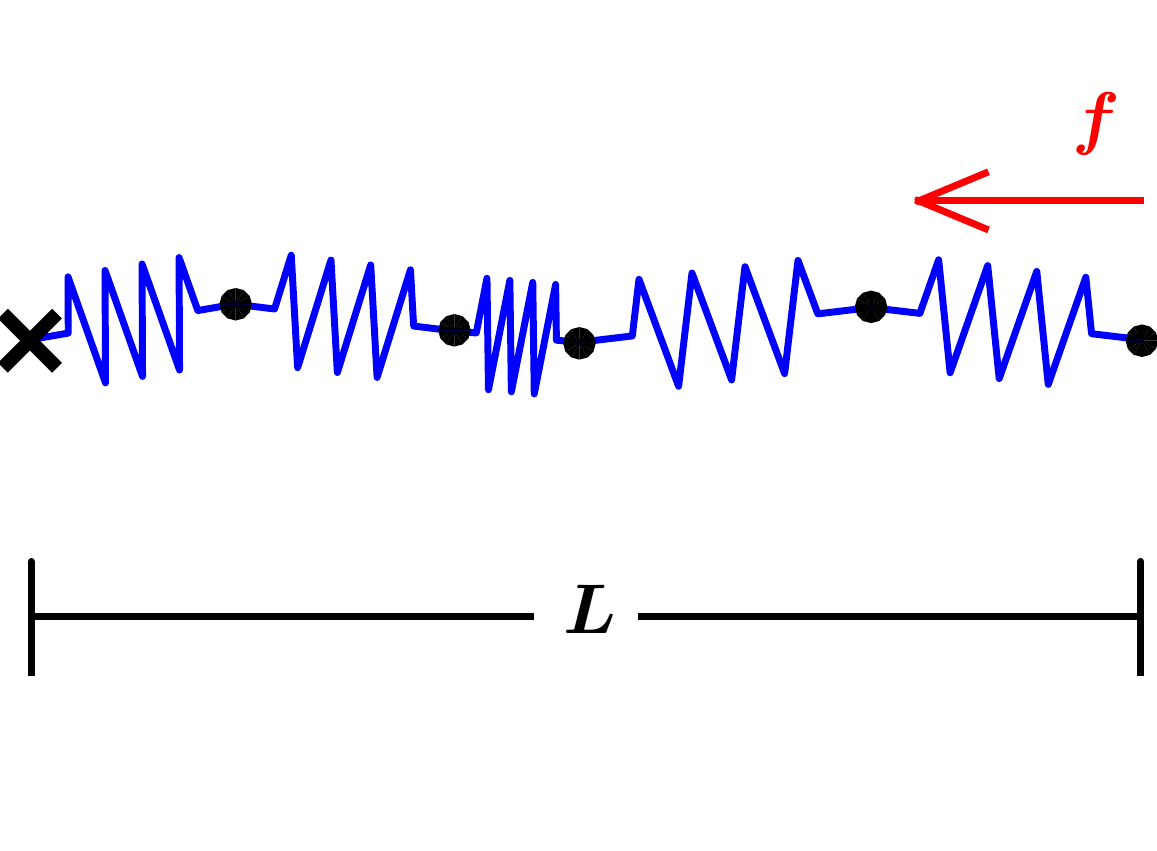}
\caption{Schematic of toy model. Each of the springs has a relaxed length of zero, and they are held at a constant extension of $L$. Representing the same system with more springs requires that each one be stiffer to maintain the same average force.}
\label{fig:toy_schematic}
\end{figure}

%\subsection{Toy Model}
The observed scaling of force fluctuations in our system can be explained with the introduction of a toy model that shares many of the features of interest. We consider a beam of mass $M$ stretched at length $L$, immersed in a heat bath. We coarse-grain it into $N$ point masses $m$ connected by $N+1$ springs of stiffness $\kappa$ (Fig.~\ref{fig:toy_schematic}), each with zero equilibrium extension. This toy model represents a simplification of that employed by others~\cite{weiner1986axial,winkler1992finite}, wherein each spring has a non-zero equilibrium extension. The beginning of the first spring is fixed at the origin, and the end of the last spring is held at the point $L\hat e_x$ in Cartesian space. In this model, all bonds are extensible, and thus the force of constraint is entirely localized to the last spring in the system, with no dependence on the velocity of the point masses. The Hamiltonian is easily separable into kinetic and potential terms:
\begin{equation}
    \mathcal{H} = \sum_{i=1}^N \frac{\|\mathbf{p}_i^2\|}{2m}  + \frac{\kappa \|\mathbf{x}_1\|^2}{2} +\sum_{i=2}^N \frac{\kappa \|\mathbf{x}_i-\mathbf{x}_{i-1}\|^2}{2}+\frac{\kappa \|\mathbf{x}_N-\mathbf{L}\|^2}{2}
\end{equation}
where each $\mathbf{p}_i$ and $\mathbf{x}_i $ represents three cartesian components of momentum or position vectors, and $\mathbf{L} = L\mathbf{e}_x$ is the displacement vector between the first and last oscillator. This separability allows us to use analytical means to derive force fluctuations. 

The total partition function for the system can be written down as $Z = \int \int \exp(-\beta \mathcal{H}) d^{3N}p\mbox{ }d^{3N}x$.
%\begin{equation}
% Z = \int_{-\infty}^{\infty}\cdots \int_{-\infty}^{\infty} e^{-\frac{\beta}{2m} (\|\mathbf{p}_1\|^2+\cdots+\|\mathbf{p}_N\|^2)} d\mathbf{p}_1 \dots d\mathbf{p}_N
% \int_{-\infty}^{\infty}\cdots \int_{-\infty}^{\infty} e^{-\frac{\beta \kappa}{2} (\|\mathbf{x}_1\|^2+\|\mathbf{x}_2-\mathbf{x}_1\|^2\cdots+\|\mathbf{x}_N-\mathbf{L}\|^2)} d\mathbf{x}_1 \dots d\mathbf{x}_N.
%\end{equation} 
All momentum integrands are Gaussians, simplifying our partition function to 
\begin{equation}
     Z = \left(\frac{2\pi m}{\beta}\right)^{3N/2}
 \int_{-\infty}^{\infty}\cdots \int_{-\infty}^{\infty} e^{-\frac{\beta \kappa}{2} (\|\mathbf{x}_1\|^2+\|\mathbf{x}_2-\mathbf{x}_1\|^2\cdots+\|\mathbf{x}_N-\mathbf{L}\|^2)} d\mathbf{x}_1 \dots d\mathbf{x}_N
\end{equation}
Each position integral can be carried out in turn on each Cartesian position, producing
\begin{equation}
Z =\frac{1}{(N+1)^{3/2}} \left(\frac{2\pi}{\beta}\sqrt{\frac{m}{\kappa}}\right)^{3N} e^{-\beta \kappa L^2/2(N+1)}
 \label{eq:z}
\end{equation}
which is identical to the partition function of a Gaussian chain\cite{winkler1992finite}. The ensemble average of the instantaneous force ($\kappa (x_N-L)$) along $\mathbf{e}_x$, is then related to the derivative of the partition function: 
%\begin{equation}
%\bar{f} = \frac{\int_{-\infty}^{\infty}\cdots \int_{-\infty}^{\infty} \kappa (x_N-L) e^{-\frac{\beta \kappa}{2} (\|\mathbf{x}_1\|^2+\|\mathbf{x}_2-\mathbf{x}_1\|^2\cdots+\|\mathbf{x}_N-\mathbf{L}\|^2)} d\mathbf{x}_1 \dots d\mathbf{x}_N}{\int_{-\infty}^{\infty}\cdots \int_{-\infty}^{\infty} e^{-\frac{\beta \kappa}{2} (\|\mathbf{x}_1\|^2+\|\mathbf{x}_2-\mathbf{x}_1\|^2\cdots+\|\mathbf{x}_N-\mathbf{L}\|^2)} d\mathbf{x}_1 \dots d\mathbf{x}_N}=-\frac{1}{\beta} \frac{\partial \log Z}{\partial L}.
%\end{equation}
\begin{equation}
\bar{f} = \frac{\int_{-\infty}^{\infty} \kappa (x_N-L) e^{-\beta U(\mathbf{x})} d^{3N}x }{\int_{-\infty}^{\infty} e^{-\beta U(\mathbf{x})} d^{3N}x}= \frac{\int_{-\infty}^{\infty} -\frac{1}{\beta} \frac{\partial }{\partial L} e^{-\beta U(\mathbf{x})} d^{3N}x }{\int_{-\infty}^{\infty} e^{-\beta U(\mathbf{x})} d^{3N}x}=-\frac{1}{\beta} \frac{d \log Z}{d L}.
\end{equation}
Hence, using Eq.~\ref{eq:z}, we obtain the generalized force conjugate to $L$
\begin{equation}
\bar{f} = \frac{\kappa}{N+1}L.
\label{eq:force1}
\end{equation}
This equation is simply Hooke's law with bulk stiffness $k=\frac{\kappa}{N+1}$, analogous to the persistence length for the wormlike chain case. Similar to Fig.~\ref{fig:force_renorm}B, increasing $N$ relieves the average stress on the system analogous to increasing the contour length.
The fluctuation can similarly be obtained by taking the second order derivative of $Z$:
\begin{equation}
\sigma_f^2=\overline{f^2}- \overline{f}^2 = -\frac{1}{\beta^2}\frac{d^2 \log Z}{d L^2} + \frac{\kappa}{\beta}
\end{equation}
which produces
\begin{equation}
\sigma_f = \sqrt{\frac{\kappa N}{\beta (N+1)}} = \sqrt{\frac{k N}{\beta}} \propto N^{0.5}
\label{eq:sigmaf}
\end{equation}
%This corresponds to Equation 12 of Weiner and Berman \cite{weiner1986axial} in the instance of $a\rightarrow 0$.
When considered at constant $\kappa$, this agrees qualitatively with the growth and saturation as a function of $N$ displayed in Fig.~\ref{fig:force_renorm}C. At fixed bulk extensibility $k$,
we also again see the phenomena where the amplitude of force fluctuations scales with degrees of freedom added to the system even when the average force is assumed not to. 
%\todo[inline]{also compare your result with \cite{winkler2010equivalence}. The result looks slightly different. }
%\subsubsection{Fluctuation and Renormalization}

%Increasing the number of springs while maintaining the same mass and bulk stiffness, and the same extension, can be thought of as a finer coarse-graining of the same system. This will cause the stiffness of individual springs to vary as $\kappa = (N+1) k$, stated above, and the mass of points to vary as $m = M /N $

What is the impact of this fluctuating force?  In statistical mechanics of many particle systems, fluctuation of intensive parameters such as force still appears to be a subject of discussion\cite{rudoi2000thermodynamic,planes2002entropic,mishin2015thermodynamic}. In our example of a single polymer chain, the fluctuating force can be given a mechanistic interpretation in terms of the actual work transmissible during a short period of time. Here, using the same Gaussian chain model above, we show that the change in energy during an adiabatic extension of the chain is bounded with respect to $N$, despite the unbounded fluctuations in the force. Imagine that the chain is allowed to extend by $\Delta$ over a time period $\tau$, shorter than the characteristic collision time between the chain and molecules in the surrounding heat bath. Using the initial microstate of the chain, we can calculate the energy difference as a result of this extension. Equations of motion in the $y$ and $z$ dimensions will be separable from those in the $x$ dimension, and will not contribute to the force. The equations of motion in the $x$ dimension for $N$ oscillators can be represented as a matrix equation
\begin{equation}
\frac{m}{\kappa}
\left(
\begin{array}{c}
     \ddot x_1 \\
     \ddot x_2 \\
     \vdots \\
     \ddot x_{N-1} \\
     \ddot x_N
\end{array}
\right) = 
-\left(
    \begin{array} {c c c c c}
    2 & -1 & 0 & \cdots & 0 \\
    -1 & 2 & \ddots & \ddots & \vdots \\
    0 & \ddots & \ddots & \ddots & 0 \\
    \vdots & \ddots & \ddots & 2 & -1 \\
    0 & \cdots & 0 & -1 & 2 
    \end{array}
    \right)
    \left(
    \begin{array}{c}
      x_1 \\
     x_2 \\
     \vdots \\
      x_{N-1} \\
      x_N
\end{array}
\right)
+ 
\left(
\begin{array}{c}
     0 \\
     0 \\
     \vdots \\
     0 \\
     -L(t)
\end{array}
\right)
\end{equation}
with a tri-diagonal matrix relating $x_i$ and $\ddot x_i$, accompanied by an inhomegenous vector representing the overall extension of the system. Solutions to this system will thus take the form 
\begin{equation}
    x_i = \sum_j^N A_{ij} \cos(\omega_j t) + B_{ij} \sin(\omega_j t) + C_i + D_i t
\end{equation}
where the summation part of the expression satisfies the homogeneous part of the equation, and the linear terms satisfy the inhomogeneous component. In the homogeneous solution, the frequencies  $\omega_j^2$ will correspond to the eigenvalues of the matrix. 
\begin{equation}
    \omega_j = 2 \sqrt{\frac{\kappa}{m}} \sin\left(\frac{1}{2}\frac{\pi j}{N+1} \right)   
\end{equation}
Note that while $\kappa$ scales as $(N+1)k$, the mass of each point will scale as $m = M/(N+1)$ to maintain a fixed linear mass density. $\omega_j$ then scales as $N+1$ when we consider finer graining of the system.
These frequencies will be unchanged by the extension of the system, which is confined to the inhomogenous part of the equation. Representing the extension by a function $L(t)$ that is piecewise linear in time, increasing uniformly from $L$ to $L+\Delta$ in the interval $t=0$ to $t=\tau$, we can obtain matching conditions for the coefficients $\mathbf{A}, \mathbf{B},\mathbf{C}$ and $ \mathbf{D}$ at times $t=0$ and $t=\tau$, when the extension begins and ends. This is done by assuming continuity of all position and momentum coordinates. Computing the difference between the Hamiltonian before and after the extension allows us to find the work done. Assuming an equilibrium distribution of energy among the normal modes, we can find the expected work and its fluctuation (see Supplemental Material for detailed calculation.). Evaluating the expectation of this difference reveals the average work
\begin{equation}
\overline{W} =  4k\Delta^2 \sum_{i=1}^N \frac{\sin^2(\omega_i \tau /2)}{\omega_i^2\tau^2}\cos^2\left(\frac{1}{2}\frac{i\pi}{N+1}\right)+\frac{k}{2} (2L\Delta+\Delta^2)
\end{equation}
Taking the lowest order terms in $\Delta$, we find $\overline W \approx k L\Delta = \Delta \overline f$ using the expression for $\overline f$ in Eq.~\ref{eq:force1}. The average value of work squared can also be found, and used to obtain the fluctuation.
\begin{equation}
\overline{ W^2}-\overline{ W}^2 =\frac{8 k\Delta^2}{\beta}\sum_{i=1}^N \frac{\sin^2(\omega_i \tau /2)}{\omega_i^2\tau^2}\cos^2\left(\frac{1}{2}\frac{i\pi}{N+1}\right)
\label{eq:workvar}
\end{equation}
If we take the limit of small $\tau$, independent of $N$, we can use the approximation $\sin(\omega_i\tau/2) \approx \omega_i \tau/2$ to reduce this to
\begin{equation}
\sigma_W^2 = \overline{ W^2}-\overline{ W}^2 \approx \frac{2k\Delta^2}{\beta}\sum_{i=1}^N \cos^2\left(\frac{i\pi}{2(N+1)}\right) = \frac{k N \Delta^2 }{\beta}
\end{equation}
which is the same result we found using equilibrium statistical mechanics (Eq.~\ref{eq:sigmaf}), albeit 
in terms of $\sigma_W =\sigma_F \times \Delta $ However, for any finite $\tau$, we will reach a limit of rescaling where
$\omega_i\tau$ is no longer negligible. This will provide a cap on our growing fluctuations, and define a scale of coarse-graining below which further fineness will not produce any change in the results. With some effort, the expression in Eq.~\ref{eq:workvar} can be manipulated to reveal
%Knowing that $\sin^2x$ is bounded by unity, a weak upper limit of the summation in Eq.~\ref{eq:workvar} is given by
%\begin{equation}
%\sigma_W^2 \le \frac{8 k\Delta^2}{\beta \tau^2}\frac{m}{\kappa} \left(\sum_{i=1}^N \csc^2\left(\frac{1}{2}\frac{i\pi}{N+1}\right) - N \right) = \frac{2 M \Delta^2}{3\beta \tau^2} \left(\frac{2N+1}{N+1} \right)
%\end{equation}
%which approaches a fixed value for large $N$. 
\begin{equation}
    \sigma_W^2 = \frac{8k\Delta^2}{\beta}\left((N+1) (2J_0(x)+\pi J_1(x)H_0(x)-\pi J_0(x)H_1(x)-\frac{1}{2} +\frac{M (J_0(x)-1))}{2(N+1) \tau^2 k} -\frac{J_1(x)}{\tau \sqrt{k/M}}\right)
\end{equation}
where $J_i$ are the Bessel functions of the first kind, and $H_i$ are the Struve functions. All of these take the same argument $x= 2 (N+1) \tau \sqrt{k/M}$. Taking the limit or large $N$, the two terms on the right will become negligible, leaving us with
\begin{equation}
    \sigma_W^2 = \frac{8k\Delta^2}{\beta}\left(\frac{\sqrt{M}}{\tau\sqrt{k}} -\frac{1}{2} \right)
\end{equation}
for $M / \tau^2k > 1/4$. The fluctuations are limited in $N$, but can increase if the time of the extension is short enough.

% What time scale are we interested in for unlooping dynamics, and how does it compare with vibrational frequencies? JTW

%some perspective on entropic force
The entropic force by a polymer is usually introduced in statistical thermodynamics by counting the number of static conformations\cite{phillips2012physical,kubo1990statistical}. Here, we used classical mechanics to reproduce the same entropic force. In this approach, the entropic force has a clear mechanistic origin from the inertial forces exerted by thermally excited constituents of the polymer, and only emerges as a fluctuation-induced quantity similar to Casimir force \cite{bartolo2002fluctuations} and depletion force\cite{bertolini2011distribution}. The kinetic origin of the entropic force had been appreciated by others\cite{neumann1980entropic,weiner1981frenkel,reineker1989deterministic,roos2014entropic}, but was only recently applied to a long polymer\cite{waters2015calculation}. When applying this approach to a looped chain, the length constraints pose an additional technical challenge in sampling chain microstates in an unbiased manner. We introduced hierarchical coordinates that allow unbiased, direct sampling of closed chains. These coordinate moves are a combination of crankshaft moves\cite{amuasi2010off,betancourt2011optimization} and concerted rotation moves\cite{bottaro2012subtle,zamuner2015efficient}. But unlike the previous numerical algorithms that applied the moves in spatially overlapping manner, we use the moves in a hierarchical manner so that they comprise generalized coordinates with well-defined partial derivatives. 

%direct simulation
Our phase-space sampling captures ``dynamic" conformations of a polymer, which is essential to access the fluctuating forces. We note that Langevin dynamics which includes the damping force cannot yield forces exerted by the chain only\cite{dolgushev2014gaussian}. In principle, dynamic conformations can be captured by molecular dynamics (MD). In one study, a hybrid MD method was used to study the dynamics of a protein-mediated DNA loop\cite{villa2005structural} by calculating the force exerted by the minimum energy conformation of the DNA and using MD to simulate the protein under this force. This procedure can be repeated to obtain relatively long-time dynamics. However, this hybrid approach does not include thermal fluctuations of the DNA loop.  We found these fluctuations to be critical to correctly determining instantaneous forces. Recently, another multiscale MD method that include the dynamics of a coarse-grained DNA loop has been introduced\cite{machado2015exploring}. It will be interesting to see whether this multiscale method can recover force fluctuation patterns similar to our prediction.  
\section{Conclusions}
Our force-sampling method offers a level of information not easily obtained from conformational statistics, with an efficiency greater than an all-atom MD simulation. Our results suggest that loop-breaking should be dominated by inertial components as opposed to a strictly elastic origin. We have demonstrated that the amplitude of fluctuations increases even as the mean goes down, and that large force values occur with a frequency greater than a Gaussian prediction. The implication of this result is that the loop stability might change with chain parameters in a way not foreseen by mean force alone. The phase space sampling method can be applied to a host of problems that involve constraints and force fluctuations. In light of growing speculation on force fluctuation as a length regulation mechanism in biology\cite{weigel2002functional,iwasa2015molecule}, we anticipate our method will prove powerful.  

\section{Acknowledgement}
We thank Kurt Wiesenfeld for helpful discussions. This work was supported by National Institutes of Health (R01GM112882). 

\bibliography{main}
\clearpage
\newcommand{\beginsupplement}{%
        \setcounter{table}{0}
        \renewcommand{\thetable}{S\arabic{table}}%
        \setcounter{figure}{0}
        \renewcommand{\thefigure}{\arabic{figure}}%
        \setcounter{equation}{0}
        \renewcommand{\theequation}{S\arabic{equation}}%
}
%\renewcommand{\figurename}{Supplementary Figure}

%\beginsupplement
%\onecolumngrid
%\section*{Supplemental Information}

%\subsection*{Computing Metric Terms}
%Rotation transformations 
%\begin{equation}
%\mathbf{r}' = \sin(\delta) \mathbf{\hat a}\times\mathbf{r} + \cos(\delta) \mathbf{r} + (1-\cos(\delta))  ( %\mathbf{\hat a}\cdot \mathbf{r})\mathbf{\hat a}
%\end{equation}

%The shift transformations can be found from four distinct rotation matrices. In the case of the forward shift transformation, these will 
%\begin{equation}
 %\mathbf{r}' = \sin(\delta) \mathbf{\hat a_{10}}\times\mathbf{r} + \cos(\delta) \mathbf{r} + (1-\cos(\delta))  (     \mathbf{\hat a_{10}}\cdot \mathbf{r})\mathbf{\hat a_{10}}
%\end{equation}
%where 
%\begin{equation}
%    \begin{array}{l c l c l}
%    \mathbf{a_{10}}& =&\mathbf{r_{100}} &\times &\mathbf{r_{101}} \\
%    \mathbf{a_1}& =&\mathbf{r_{10}} &\times &\mathbf{r_{11}} 
%    \end{array}
%\end{equation}

\end{document}